\documentclass[]{JHEP3} 

\usepackage{epsfig,multicol}

\newcommand\fverb{\setbox\pippobox=\hbox\bgroup\verb}
\newcommand\fverbdo{\egroup\medskip\noindent%
                        \fbox{\unhbox\pippobox}\ }
\newcommand\fverbit{\egroup\item[\fbox{\unhbox\pippobox}]}
\newbox\pippobox

\def\as{\alpha_s}
\def\aew{\alpha_{\rm ew}}
\newcommand{\cO}[1]{\mathcal{O}\left(#1\right)}
\def\slsh{\rlap{$\;\!\!\not$}}     
\def\cG{c_\Gamma}

\def\ETmiss{\slash \!\!\!\!\!\!\ E_{\rm T}}
\def\pTmiss{\slash \!\!\!\!\!\!\ \vec p_{\rm T}}

\def\MSbar{\overline{\mbox{\small MS}}}

\def\bentarrow{\:\raisebox{1.3ex}{\rlap{$\vert$}}\!\rightarrow}
\def\dkp#1#2#3#4{
\begin{array}{r c l}
#1 & \rightarrow & #2#3 \\
 & & \phantom{\; #2}\bentarrow #4
\end{array}}
\def\bothdk#1#2#3#4#5{
\begin{array}{r c l}
#1 & \rightarrow & #2#3 \\
 & & \:\raisebox{1.3ex}{\rlap{$\vert$}}\raisebox{-0.5ex}{$\vert$}%
\phantom{#2}\!\bentarrow #4 \\
 & & \bentarrow #5
\end{array}
}
\def\cA{{\cal A}}
\def\nn{\nonumber}

\title{Next-to-leading order predictions for WW + jet distributions at
the LHC}

\author{John M. Campbell, \\
Department of Physics and Astronomy,\\
University of Glasgow, Glasgow G12 8QQ, UK\\
Email:~\email{j.campbell@physics.gla.ac.uk}}
\author{R. Keith Ellis, \\ 
Fermilab, Batavia, IL 60510, USA\\
Email:~\email{ellis@fnal.gov}}
\author{Giulia Zanderighi\\
The Rudolf Peierls Centre for Theoretical Physics\\  
Department of Physics, University of Oxford \\
1 Keble Road, OX1 3NP, Oxford, UK \\
Email: ~\email{g.zanderighi1@physics.ox.ac.uk}
}
\preprint{\hepph{yymm.xxxx}\\
\today\\
CERN-PH-TH-07/181\\
FERMILAB-PUB-07-529-T\\
OUTP-07-11-P}

\abstract{ We present numerical results for the production of a
$W^+W^-$ pair in association with a jet at the LHC in QCD at
next-to-leading order (NLO).  We include effects of the decay of the
massive vector bosons into leptons with spin correlations and
contributions from the third generation of massive quarks.  The
calculation is performed using a semi-numerical method for the virtual
corrections, and is implemented in MCFM.  In addition to its
importance {\it per se} as a test of the Standard Model, this process
is an important background to searches for the Higgs boson and to many
new physics searches. As an example, we study the impact of NLO
corrections to $W^+W^-+$~jet production on the search for a Higgs
boson at the LHC.  }


\begin{document} 


\section{Introduction}

The search for the Higgs boson at the Large Hadron Collider (LHC) will
rely on analysing many types of events related to its different
production and decay modes~\cite{AtlasTDR,CMSPhysics}.  If a Standard
Model (SM) Higgs boson mass lies in the range $155 <m_H< 185 $~GeV,
the production of a Higgs boson that decays to $W$ pairs, $H \to W^+
W^-$, is expected to be a significant channel. It could even be the
discovery mode for the Higgs boson, particularly if the mass of the
Higgs boson lies very close to the threshold for the production of $W$
pairs, $M_H=2 M_W$.  Note that the $W$-bosons can both be real ($M_H
\geq 2 M_W$) or virtual ($M_H < 2 M_W$).

\begin{figure}[t]
\begin{center}
\includegraphics[angle=0,scale=0.7]{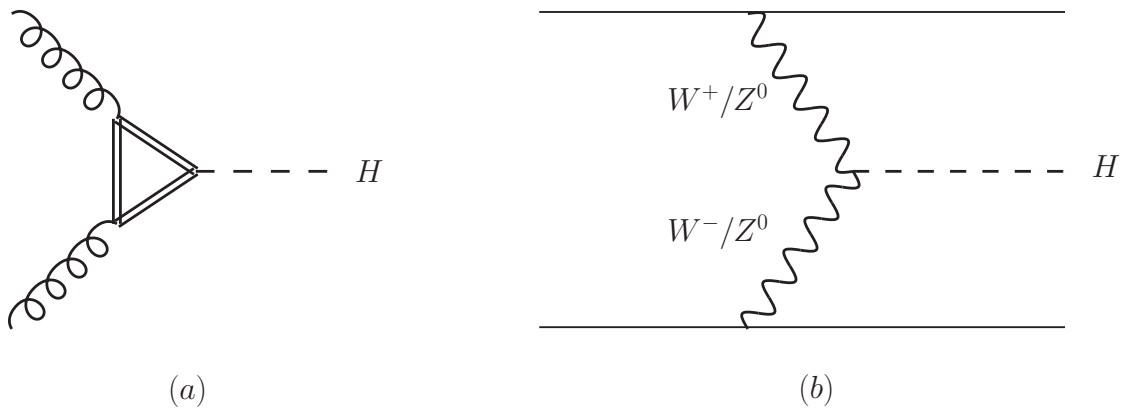}
\end{center}
\caption{The two primary mechanisms for producing a Higgs boson in association
with jets at the LHC.}
\label{fig:hprod}
\end{figure}

There are two main mechanisms for producing a Higgs boson that leads
to such final states at the LHC. The Higgs boson can be produced as
the result of gluon fusion, with the Higgs boson coupling to an
intermediate heavy quark loop (Fig.~\ref{fig:hprod}(a)), with further
QCD radiation leading to additional jets being observed in the final
state.
Alternatively, the Higgs boson may be produced with a sizeable cross
section via weak boson fusion (WBF - see Fig.~\ref{fig:hprod}(b)), in
which the decay products of the Higgs boson are naturally accompanied
by two forward jets. Therefore the channel $W^+ W^-$ accompanied by
$0$, $1$ or $2$ jets will be subject to intense scrutiny at the LHC
and all SM backgrounds should be investigated as fully as possible.

If such a Higgs boson exists, the $WW$ + 0 jet sample is expected to
contain events originating from a Higgs boson that is predominantly
produced via gluon fusion, with the largest background being the
normal QCD production of $WW$ pairs.  This signal has been calculated
not only to NLO~\cite{Dawson:1990zj,Djouadi:1991tk} but also up to
NNLO~\cite{Harlander:2000mg,Anastasiou:2002yz}, whilst the NLO
corrections to this background are also
known~\cite{Ohnemus:1991kk,Campbell:1999ah,Dixon:1999di}.

If instead the sample is selected by requiring two forward jets then
the putative Higgs boson is preferentially produced via the WBF
mechanism.  This is now expected to be the single most significant
channel in the range $130 < m_H < 190 $~GeV (see for example
Ref.~\cite{Asai:2004ws}).  In this case, the signal has been known to
NLO for some time~\cite{Han:1992hr,Figy:2003nv,Berger:2004pc}, whilst
the NLO calculation of the dominant background from the production of
$WW+2$~jets is currently unknown.

It has recently been suggested that the Higgs signal significance can
be improved by considering also the final state which consists of the
leptonic decays of $W$-pairs, plus at least one additional
jet~\cite{Quayle}. In this study, only a single jet is demanded at
large rapidity. This has the effect of substantially reducing the
backgrounds whilst not unduly compromising the signal because of the
nature of the WBF events. This renders such a search feasible, albeit
with significant SM backgrounds whose size should clearly be assessed
beyond the LO approximation. Dominant among these is the production of
$WW$+jet, the NLO corrections to which we present in this paper. Since
the final state is identified through the leptonic decay modes of the
$W^+W^-$ system to $\ell^+ \ell^-$ and missing energy, we include in
our calculation the leptonic decays of the $W$'s and their spin
correlations~\footnote{ In the final stages of preparation of this paper, an
independent calculation of this process has been
presented~\cite{Dittmaier:2007th}.  The results of that calculation
appear consistent with the ones indicated by this paper, although
Ref.~\cite{Dittmaier:2007th} does not contain an evaluation of the NLO
corrections at a specific phase space point that we could exactly
compare with in a straightforward way.  We note however that the
calculation of Ref.~\cite{Dittmaier:2007th} does not include the decay
products of the $W$ bosons and their correlations.}.
The calculation is performed using a recently-developed semi-numerical
method for computing virtual 1-loop
corrections~\cite{Giele:2004iy,Giele:2004ub,Ellis:2005zh}.

The importance of this calculation has already been recognized by its
appearance on the list of next-to-leading order priorities assembled
at the Les Houches workshop in 2005~\cite{Buttar:2006zd}. This is
motivated largely by its obvious importance to the search for the
Higgs boson and for other signals of new physics at the
LHC. Nevertheless, the $WW$+jet process is interesting in its own
right as a test of our theoretical understanding of the SM. In the
same way that the SM has been probed experimentally at the Tevatron by
measurements of the $WW$ and $W+$~jets cross sections~(for recent
examples, see Refs.~\cite{Abazov:2006xq} and~\cite{Affolder:2000mz}),
the rate of events at the LHC will allow similar studies of $WW$+jet
events at even higher energy scales.  In addition, this calculation is
a necessary stepping stone to highly-desired NLO calculations of $2
\to 4$ processes such as the aforementioned $WW+2$~jet
process~\cite{Buttar:2006zd}, as well as $W+3$~jet production.

The structure of the paper is as follows. In
section~\ref{sec:calculation} we discuss the overall structure of the
calculation, with attention paid to the finite contribution arising
from internal loops of massive quarks.  Section~\ref{sec:results}
presents the results of our calculation with two types of event
selection, one indicative of a typical inclusive analysis at the LHC
and the other relevant for a search for the Higgs boson. To that end,
we perform a parton-level analysis of all the major background
processes using the general purpose NLO program
MCFM~\cite{Campbell:1999ah}. Our conclusions are contained in
section~\ref{sec:conclusions}. Finally, the appendices consist of
analytical results for some of the massive triangle diagrams that
appear in our calculation, together with explicit numerical results
for the 1-loop corrections at a particular phase space point.

\section{Structure of the calculation} \label{sec:calculation}
\subsection{Lowest order}
\begin{figure}[t]
\begin{center}
\includegraphics[angle=270,scale=0.7]{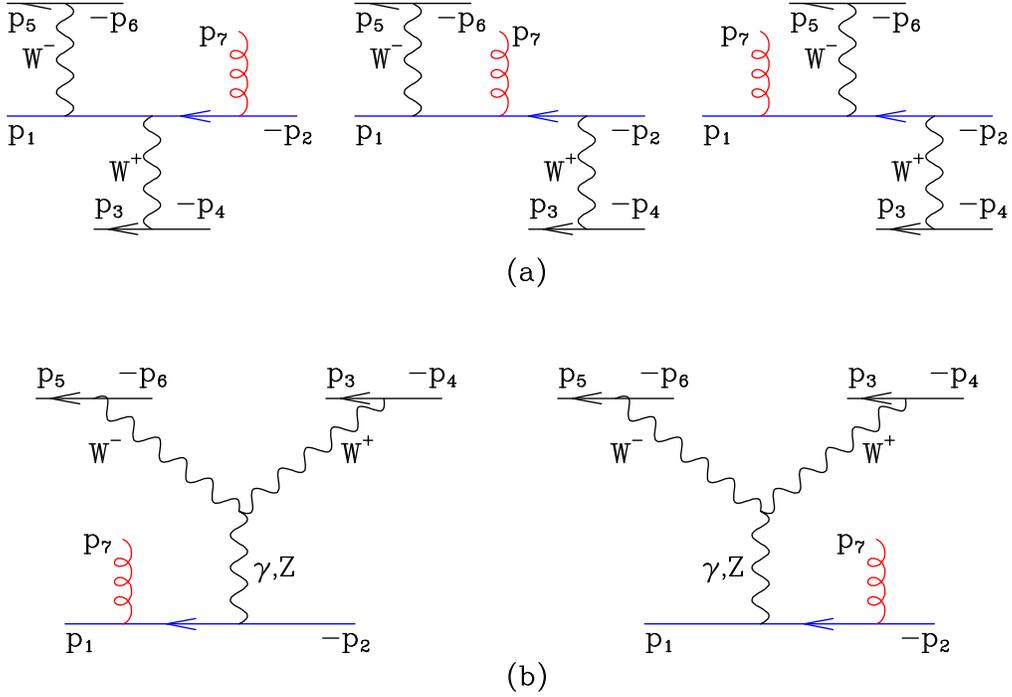}
\end{center}
\caption{Leading order diagrams for up quark annihilation to $W^+W^-+g$.}
\label{WpWmg}
\end{figure}

Let us consider as the lowest order, $\cO{\aew^4 \as}$, reference
process the one in which the up quarks annihilate into a $W^+ W^- $
pair,
\begin{equation}
 0 \to u(p_1) +\bar{u}(p_2)  + W^+(\ell(p_3)+\bar{\ell}(p_4)) 
+ W^-(\ell(p_5)+\bar{\ell}(p_6))+g(p_7)\,.
\end{equation}
The Feynman diagrams and the momentum assignments are shown in
Fig.~\ref{WpWmg}. All momenta are outgoing.
The results for a positive helicity gluon and left handed quark line
are determined in terms of the two primitive amplitudes given
below~\cite{Dixon:1998py}~\footnote{Note that apart from the overall
sign,
 this agrees with the result obtained in~\cite{Dixon:1998py} Eqs.(2.22)
  and (2.23) for $0 \to u(p_1) +\bar{u}(p_2)  + W^-(\ell(p_3)+\bar{\ell}(p_4)) 
+ W^+(\bar \ell(p_5)+\ell(p_6))+g(p_7)$.},
\begin{eqnarray}
\label{A_7a}
A_7^{(a)}&=&
 \frac{\langle 15 \rangle } {\langle 17 \rangle s_{56}s_{34} s_{156}} 
 \Bigg[\frac{\langle 15 \rangle [56] [42] \langle 3|2+4|7\rangle }{s_{234}}
+\frac{\langle 3|1+5|6 \rangle 
               \langle 1|2+7|4 \rangle} {\langle 27 \rangle}
\Bigg]\,,
\end{eqnarray}
\begin{eqnarray}
\label{A_7b}
A_7^{(b)}&=&
  \frac{1}{\langle 17\rangle \langle 27\rangle s_{56}s_{34} s_{127}}
 \Bigg[\langle 15\rangle \langle 1 |2+7|6 \rangle \langle 3 |5+6|4 \rangle 
\nonumber \\
   &-&\langle 13\rangle \langle 1 |2+7|4 \rangle \langle 5 |3+4|6 \rangle 
-\langle 35\rangle [46] \langle 1 |(3+4)(2+7)|1 \rangle 
\Bigg]\,,
\end{eqnarray}
where $s_{ij}=(p_i+p_j)^2,s_{ijk}=(p_i+p_j+p_k)^2$.  $\langle
ij\rangle$ and $[ij]$ are the standard spinor products for massless
vectors such that $\langle ij\rangle [ji]=s_{ij}$.  The results for
other helicities may be obtained by permutations of the results in
Eqs.~(\ref{A_7a}, \ref{A_7b}).
Since we are interested in $W^+W^-$+jet as a background to Higgs boson
or new physics searches, we need to go beyond the zero-width
approximation and in general we have $q_V^2 \ne M_V^2$. We introduce
the $W$ propagators with a Breit-Wigner form,
\begin{equation}
P_W(s) \equiv \frac{s}{s-M_W^2+i \Gamma_W M_W}\,. 
\end{equation}
If needed one can also introduce a width for the $Z$-boson.  To do
this in a gauge invariant (though not unique) way one can multiply the
whole amplitude by an additional
factor~\cite{Baur:1991pp,Kurihara:1994fz},
\begin{equation}
P_Z(s) \equiv \frac{s-M_Z^2}{s-M_Z^2+i \Gamma_Z M_Z}\,.
\end{equation}

To obtain the full amplitude the results of Eqs.(\ref{A_7a},
\ref{A_7b}) must be dressed with appropriate couplings,
\begin{eqnarray}
\label{eq:treeAu}
&&\cA^{A}(u^L_1,\bar{u}^R_2,\nu_3,e^+_4,\mu^-_5,\bar{\nu}_6,g_7^R)= 
\nonumber\\ &&
\sqrt{2}\, t^A _{i_1 i_2}\, H
\Bigg[A_7^{(a)}(1,2,3,4,5,6,7)+C_{L,\{u\}}(s_{127}) A_7^{(b)}(1,2,3,4,5,6,7)\Bigg]\nonumber,\\ 
&&
\cA^A(u^R_1,\bar{u}^L_2,\nu_3,e^+_4,\mu^-_5,\bar{\nu}_6,g_7^R)
=    \sqrt{2}\, 
t^A _{i_1 i_2}\, H
\Bigg[C_{R,\{u\}}(s_{127}) A_7^{(b)}(2,1,3,4,5,6,7)\Bigg] \, ,
\end{eqnarray}
where the couplings $C_{L,\{u\}}$ and $C_{R,\{u\}}$ that appear will be defined
shortly and $H$ is an overall factor given by,
\begin{equation}
H= i g_s g_W^4 P_W(s_{34})P_W(s_{56})P_Z(s_{127})\,.
\label{Overall}
\end{equation}
The corresponding results for 
processes involving $d$ quarks can be written as follows,
\begin{eqnarray}
\label{eq:treeAd}
&&\cA^A(d^L_1,\bar{d}^R_2,\nu_3,e^+_4,\mu^-_5,\bar{\nu}_6,g_7^R)
= 
\nonumber\\&& \sqrt{2}\, t^A_{i_1 i_2}\, H
\Bigg[A_7^{(a)}(1,2,5,6,3,4,7)+C_{L,\{d\}}(s_{127}) A_7^{(b)}(1,2,5,6,3,4,7)\Bigg]\nonumber, \\
&&
\cA^A_d(d^R_1,\bar{d}^L_2,\nu_3,e^+_4,\mu^-_5,\bar{\nu}_6,g_7^R)
= \sqrt{2}\, t^A _{i_1 i_2}\,  H
\Bigg[C_{R,\{d\}}(s_{127}) A_7^{(b)}(2,1,5,6,3,4,7)\Bigg],
\end{eqnarray}
where we now specify the left- and right-handed couplings of up- and down-type
quarks as,
\begin{eqnarray}
C_{L,\{ {u\atop d} \}}
 (s) &=& \pm \Bigg [2 Q \sin^2 \theta_W 
+\frac{s (2 T_3^f - 2 Q \sin^2 \theta_W)}{s-M_Z^2}\Bigg]\,, \\
C_{R,\{ {u\atop d} \}}(s) &=& \pm 2 Q \sin^2 \theta_W 
\Bigg[ 1- \frac{s}{s-M_Z^2}\Bigg]\, .
\end{eqnarray}
In these formulae, $T_3^f=\pm \frac{1}{2}$ and $t^A_{ij}$ is the
colour matrix in the fundamental representation for the emitted gluon,
normalized so that Tr$(t^A t^B) =\frac{1}{2}\, \delta^{AB}$.
$g_s$ represents the strong coupling and the weak coupling is related
to the Fermi constant by $g_W^2/(8 M_W^2)=G_F/\sqrt{2}$.  The extra
sign for the $d$-type quarks is due to the fact that the triple boson
term does not require the shift ($3 \leftrightarrow 5,4
\leftrightarrow 6$) when we go from $u$ to $d$. If it is included anyway then
an extra minus sign results.

The leading order cross-section is obtained by considering all
possible crossings of the quarks and gluon and by summing the squared
amplitudes for all final state helicities and averaging over the
spin and colour of the initial state partons. 
At leading order the cross-section receives contributions from
processes with an incoming quark-antiquark pair or with one
(anti-)quark and one gluon in the initial state.

\subsection{Real corrections}
The real matrix element corrections to $W^+W^-+1$~jet production are
obtained by including all crossings of results for the two basic
processes,
\begin{eqnarray}
 0 &&\to u(p_1) +\bar{u}(p_2)  + W^+(\ell(p_3)+\bar{\ell}(p_4)) 
+ W^-(\ell(p_5)+\bar{\ell}(p_6))+g(p_7)+g(p_8)\,,\nonumber \\
 0 &&\to u(p_1) +\bar{u}(p_2)  + W^+(\ell(p_3)+\bar{\ell}(p_4)) 
+ W^-(\ell(p_5)+\bar{\ell}(p_6))+q'(p_7)+\bar q'(p_8)\,.\nonumber \\
\end{eqnarray}
These tree-level matrix elements have been checked
against the results of Madgraph~\cite{Maltoni:2002qb}.
At $\cO{\aew^4 \as^2}$ the cross-section receives contributions from
processes with two incoming gluons and a $qq$ ($\bar q \bar q$)-pair
as well.
Soft and collinear singularities are handled using the dipole
subtraction scheme~\cite{Catani:1997vz}.

\subsection{Virtual corrections}

\begin{figure}[t]
\begin{center}
\includegraphics[angle=270,scale=0.7]{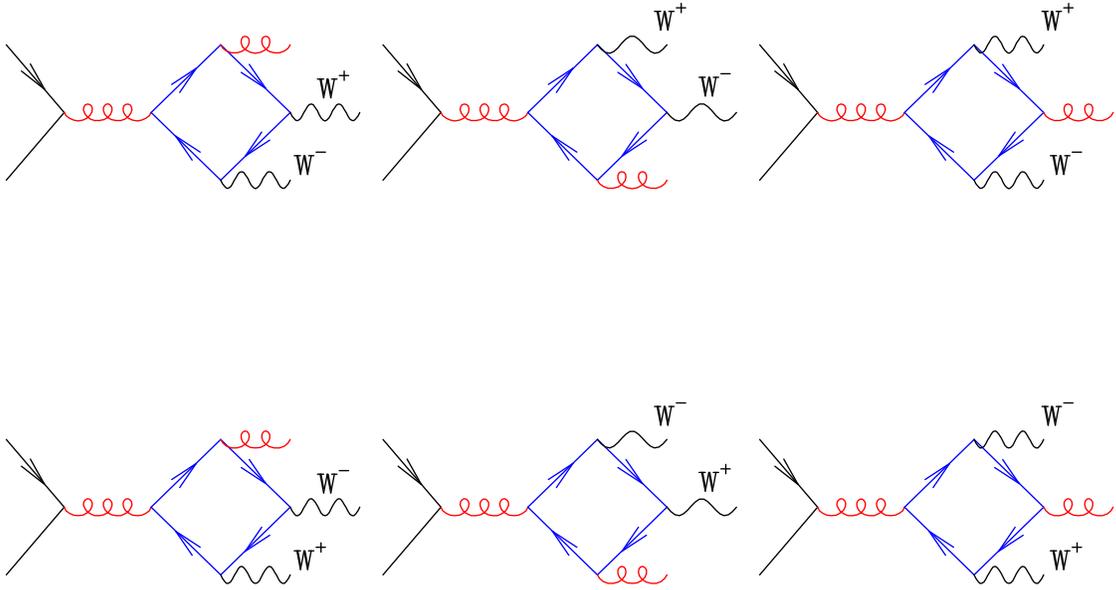}
\end{center}
\caption{Fermion loop diagrams with $W^+W^-$ contributing at one loop.}
\label{box}
\end{figure}

For the virtual corrections we have to consider one-loop QCD
corrections to the tree-level processes in figure~\ref{WpWmg}. We have
altogether 30 diagrams where the $W$-bosons are attached directly to
the incoming quark lines and which result from dressing the diagrams
of figure
\ref{WpWmg}(a) with virtual gluons. 
In addition, we have the 6 diagrams of Fig.~\ref{box} which have two
charged bosons attached to a fermion loop.

\begin{figure}[t]
\begin{center}
\includegraphics[angle=270,scale=0.6]{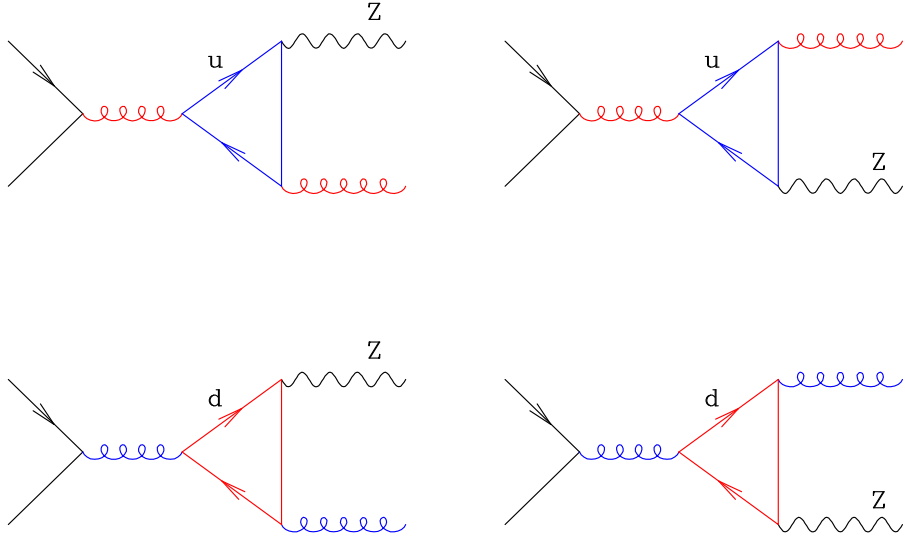}
\end{center}
\caption{Fermion loop diagrams with $Z$ contributing at one loop.}
\label{triangle}
\end{figure}

There are 11 diagrams which result from dressing the diagrams of
Fig.~\ref{WpWmg}b with virtual gluons. In addition there are the four
diagrams of Fig.~\ref{triangle} which have a $Z$ boson attached to a
fermion loop and 6 bubble diagrams with a fermion loop which vanish
trivially because of colour conservation.
The full one-loop virtual corrections we consider here involve
fermionic corrections coming from two massless families circulating in
the closed fermion loop and from a third massive generation, $m_b \neq
m_t \ne
 0$. (We could as well choose to set $m_b \to 0$ and keep only $m_t$
 finite).

The massless contribution to the virtual matrix elements have been
computed using the semi-numerical approach presented in
Refs.~\cite{Giele:2004iy,Giele:2004ub,Ellis:2005zh} based on
Davydychev reduction of tensor integrals followed by scalar reduction
of higher dimensional scalar integrals.
Results were checked using a second method based on Passarino-Veltman
reduction \cite{Passarino:1978jh}.  The method is considerably more
efficient than the one used
for~\cite{Ellis:2005qe,Campbell:2006xx,Ellis:2006ss} since we have now
a booking of computed integrals not only for higher dimensional scalar
integrals, but also for tensor ones.

An additional complication compared to the applications of the
semi-numerical method considered before is that we have closed fermion
traces involving one or two $\gamma_5$ matrices. We adopt the
t'Hooft-Veltman prescription and split $\gamma_\mu$ matrices into a
four-dimensional part $\hat \gamma_{\mu}$ which anti-commutes with
$\gamma_5$ and an $\epsilon$-dimensional part $\tilde \gamma_\mu$
which commutes with $\gamma_5$,
\begin{equation}
\gamma_\mu = \hat \gamma_{\mu} + \tilde \gamma_{\mu}\,, \qquad
\{\hat \gamma_\mu, \gamma_5\} = 0\,, \qquad 
[\tilde \gamma_\mu, \gamma_5] = 0\,.
\end{equation}

For the massive contributions we have to consider box diagrams with
two $W$ bosons attached to the loop, shown in Fig.~\ref{box} and
triangle diagrams where a $Z$ boson is attached to the quark loop
shown in Fig.~\ref{triangle}.  We computed the triangle diagrams both
fully analytically (results are given in App.~\ref{app:triangles}) and
numerically using LoopTools~\cite{Hahn:2000jm}. Box integrals have
been computed only numerically. Results for these amplitudes, as well
as for the massless amplitudes, are given in Appendix~\ref{app:numres}
for one randomly chosen phase space point.

The whole calculation is incorporated into a private version of the
general-purpose next-to-leading order code
MCFM~\cite{Campbell:1999ah}. The code is flexible enough to allow
running with on-shell or off-shell bosons.

\section{Results}
\label{sec:results}

In the following we present for results for $W^+W^-$ + jet production
at the LHC ($pp$ collisions, $\sqrt{s} = 14$ TeV). We use the
following electroweak parameters:

\begin{displaymath}
\begin{array}{lll}
M_Z = 91.188 \mbox{ GeV}\,,\qquad  &M_W = 80.419 \mbox{ GeV}\,, &M_H = 170 \mbox{ GeV}\,, \\
\Gamma_Z = 2.49 \mbox{ GeV}\,,\qquad &\Gamma_W = 2.06 \mbox{ GeV}\,,\qquad &G_F= 1.16639 10^{-5}\,,  
\label{eq:input}
\end{array}
\end{displaymath}
and furthermore we have,
\begin{equation}
m_t = 172.5 \mbox{ GeV}\,,\qquad \Gamma_t = 1.48 \mbox{ GeV}\,, \qquad m_b = 4.62 \mbox{ GeV}\,. 
\label{eq:qcdinput}
\end{equation}
For the calculation of $WW+$jet at NLO (LO) we use MRST2004f4nlo
(MRST2004f4lo) parton distribution functions corresponding to
$\alpha_s(M_Z)=0.1137$ ($\alpha_s(M_Z)=0.1251$)~\cite{Martin:2006qz}.
Furthermore we use a four-flavour running and four-flavour evolution
of pdfs and the sum over partons in the incoming protons runs over
light partons only ($d,u,s,c$).
When we consider the Higgs search in section~\ref{sec:higgssearch} we
use standard, five-flavour parton densities with five-flavour running
of the coupling for the signal processes and all backgrounds other
than $WW$+jet.  Specifically, we use MRST2004~\cite{Martin:2004ir}.
To define a jet we run the inclusive $k_T$ algorithm with $R=0.6$.  In
the following we consider two sets of cuts, a more inclusive one and
one designed to suppress QCD $W^+W^-$ + 1 jet events and other
backgrounds compared to $H(\to W^+W^-)$ + 1 jet.

\subsection{Inclusive cuts} We consider first a fairly minimal set of
cuts to examine the effect of the NLO corrections to the $WW$+jet
process in an inclusive study. We require at least one jet with,
\begin{equation}
P_{\rm t,j1} > 30 \mbox{ GeV}, \qquad |\eta_{\rm j1}| <4.5 \, ,
\label{eq:incljetcuts}
\end{equation}
and then some basic requirements on the decay products of the $W$'s: a
minimum missing transverse energy and two opposite sign charged
leptons with,
\begin{equation}
P_{\rm t, miss} > 30\mbox{ GeV}\,,\qquad  
P_{\rm t, \ell_1} > 20 \mbox{ GeV}\,,\qquad  
P_{\rm t, \ell_2} > 10 \mbox{ GeV}\,,\qquad   
|\eta_{\rm \ell_1 (\ell_2)}|< 2.5 \,.  
\label{eq:inclleptoncuts}
\end{equation}
Furthermore we require the leptons to be isolated, i.e. we impose,
\begin{equation}
R_{\rm j, \ell_1} > 0.4\,,\qquad 
R_{\rm j, \ell_1} > 0.4\,,\qquad 
R_{\rm \ell_1, \ell_2} > 0.2\,. 
\label{eq:inclisolation}
\end{equation}
Together, we refer to the cuts in equations~(\ref{eq:incljetcuts}),
~(\ref{eq:inclleptoncuts}) and~(\ref{eq:inclisolation}) collectively
as ``cuts I''.

\begin{figure}
\begin{minipage}{.9\textwidth}
  \centering
  \includegraphics[angle=270,width=.7\textwidth]{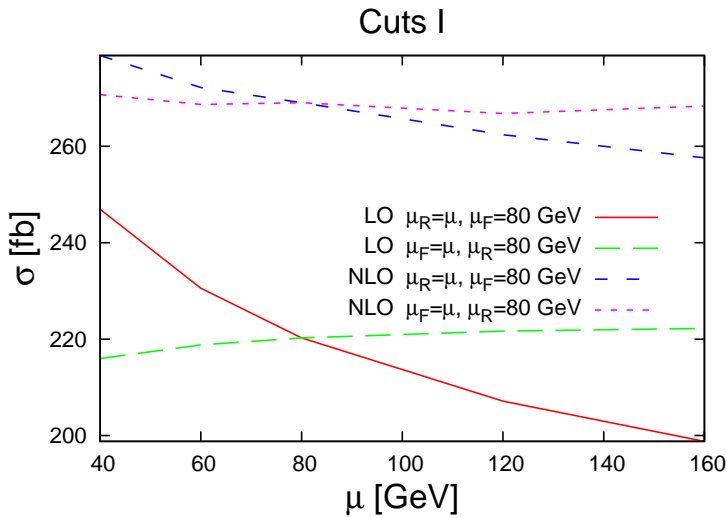}
\end{minipage} \caption{Scale dependence for the cross section with
cuts I only, as specified in the text.}
\label{fig:scaledepincl}
\end{figure}

In figure~\ref{fig:scaledepincl} we plot the total cross section with
cuts I.
We fix the renormalization scale (or the factorization scale) to be
equal to $80$~GeV and vary the other scale around this value.  We see
that the NLO correction is sizeable, of the order of $25$\%, and that
this is not covered by the scale variation of the LO prediction.
While the factorization scale dependence is quite mild both at LO and
at NLO, the dominant variation is due to the renormalization scale.
This dependence is reduced by roughly a factor two at NLO.

Since our calculation is implemented in the parton-level Monte Carlo
program MCFM, we are also able to examine the effect of the QCD
corrections on any infrared-safe observable.
\begin{figure}
\begin{minipage}{.48\textwidth}
\centering 
\includegraphics[angle=270,width=1.\textwidth]{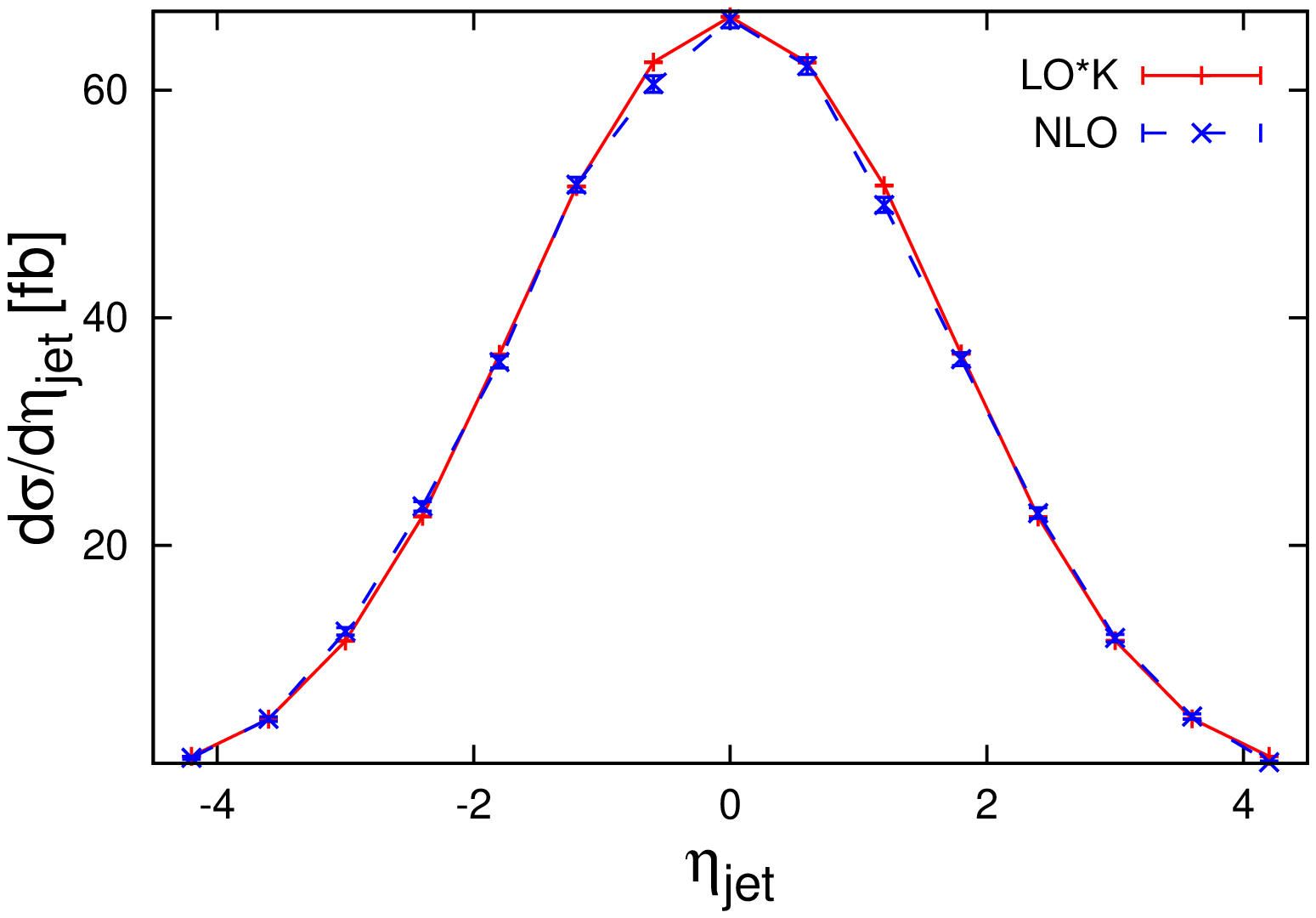}
\end{minipage}
\begin{minipage}{.48\textwidth}
\centering 
\includegraphics[angle=270,width=1.\textwidth]{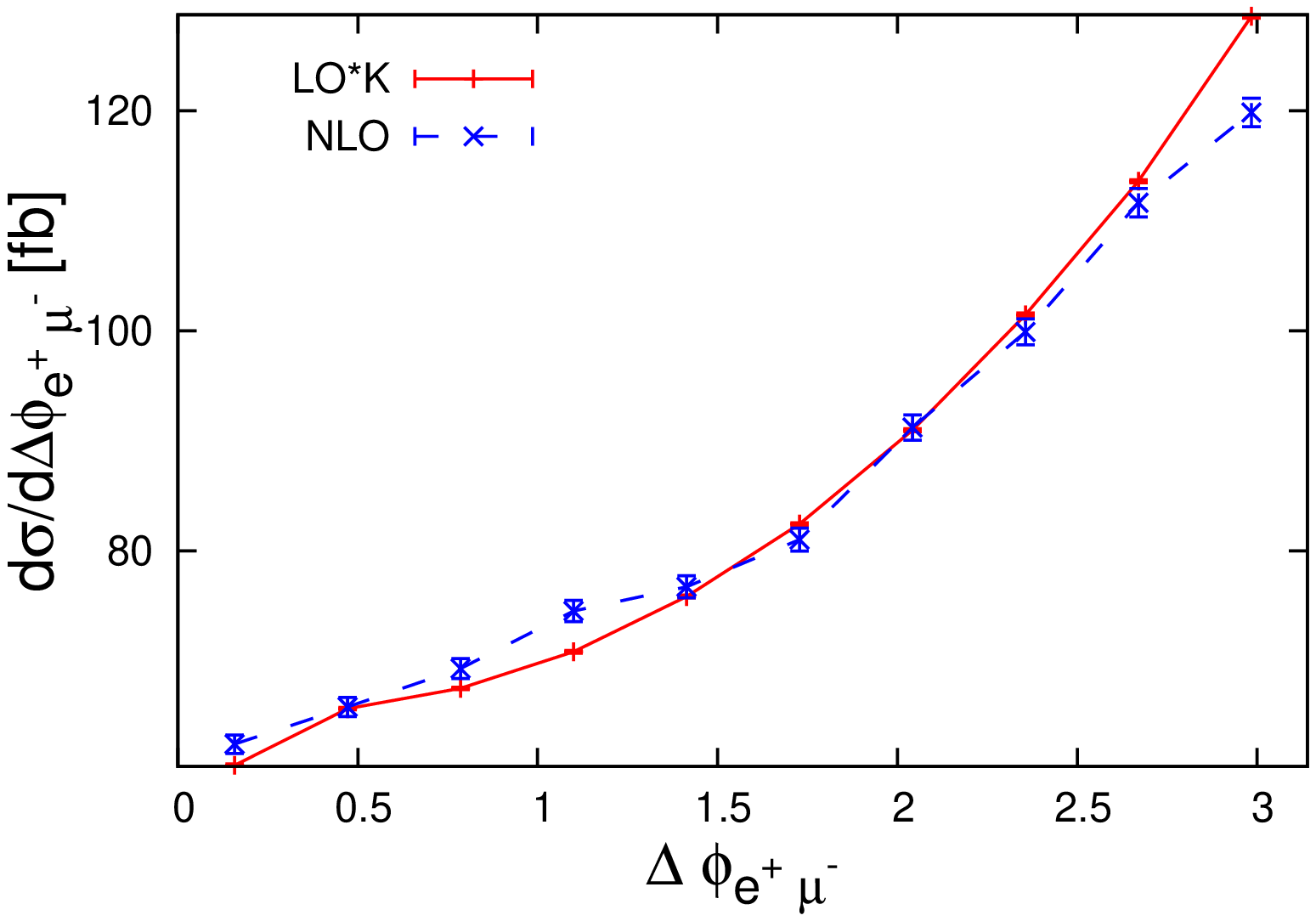}
\end{minipage} \\
\begin{minipage}{.48\textwidth}
\centering 
\includegraphics[angle=270,width=1.\textwidth]{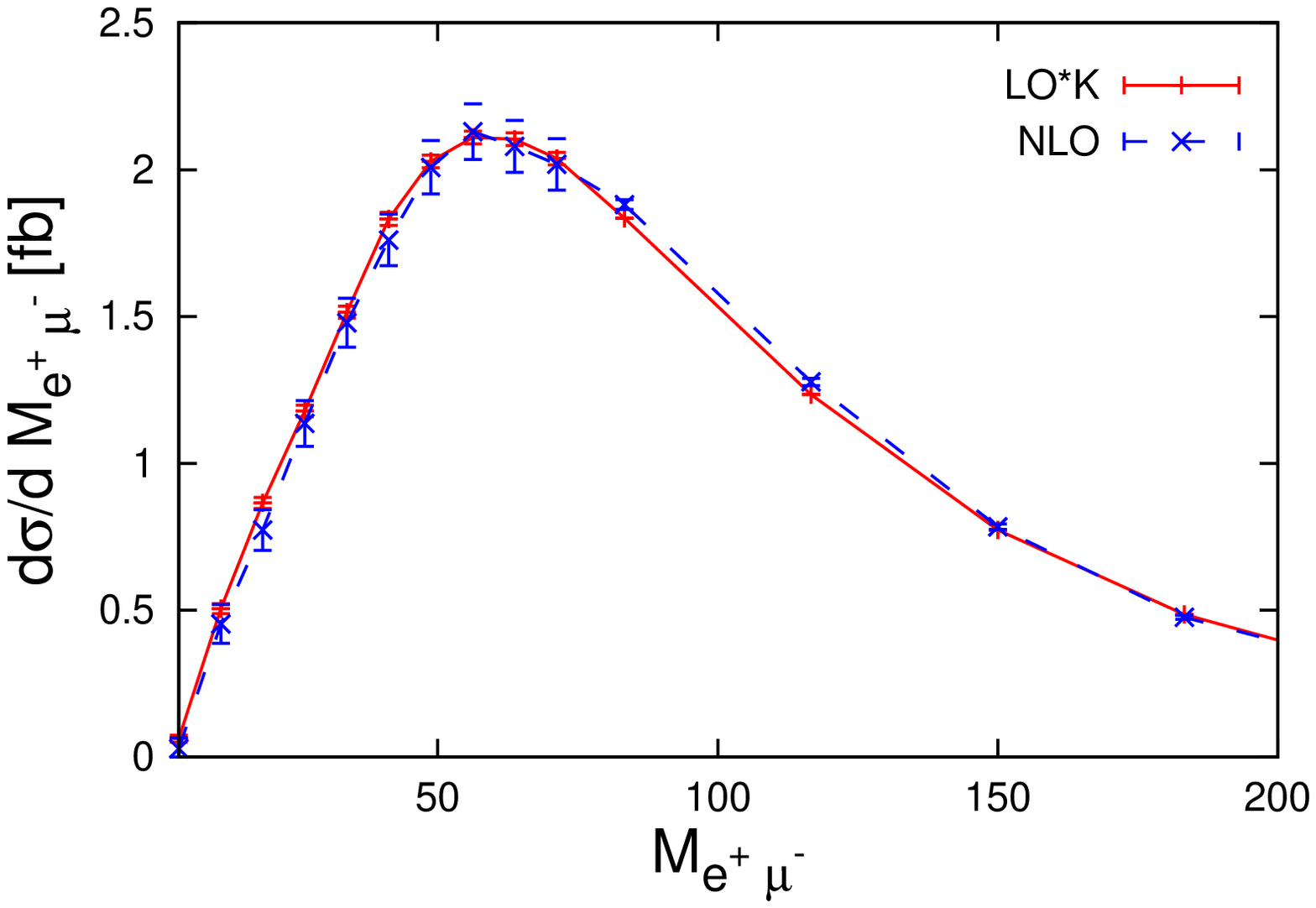}
\end{minipage}
\begin{minipage}{.48\textwidth}
\centering 
\includegraphics[angle=270,width=1.\textwidth]{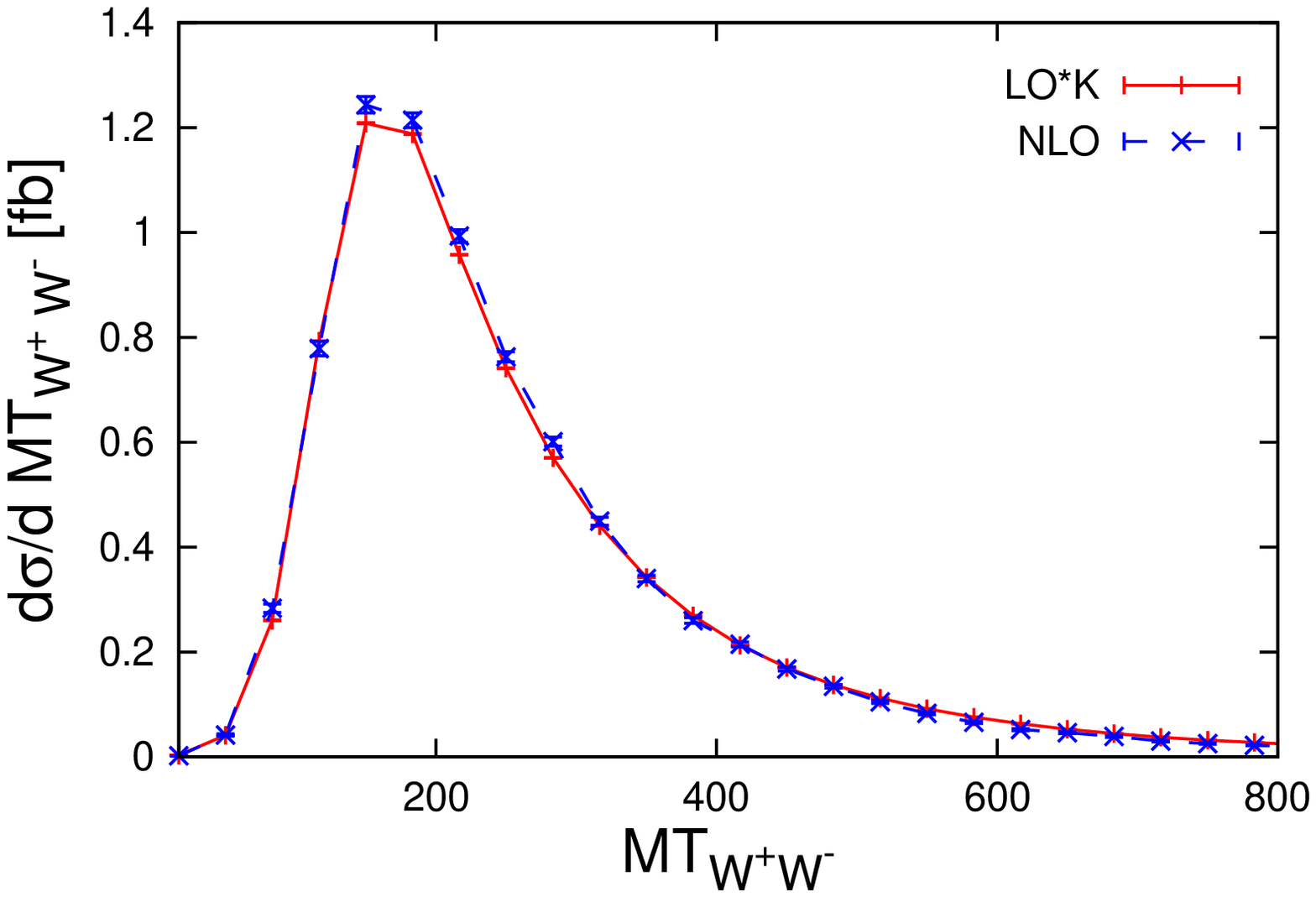}
\end{minipage}
\caption{A comparison of the shapes of various kinematical distributions in
the $WW$+jet process at LO and NLO with cuts I. The observable $M_{\rm
T, WW}$, the transverse mass of the $W^+ W^-$ pair, is defined by
equations~(\protect\ref{eq:mtww1}) and~(\protect\ref{eq:mtww2}).
\label{fig:inclplots}}
\end{figure}
In Figure~\ref{fig:inclplots} we compare the LO and NLO predictions at
$\mu_F=\mu_R=80$~GeV for a few examples of distributions that are
often important in experimental analyses: the rapidity $\eta_{\rm
jet}$ of the leading jet, the azimuthal angle $\Delta \phi_{e\mu}$
between the two charged leptons, the invariant mass of the two charged
lepton system $M_{e\mu}$ and the transverse mass of the $W^+W^-$ pair
defined as in Ref.~\cite{Rainwater:1999sd},
\begin{equation}
M_{\rm T, WW} = \sqrt{(\ETmiss +E_{T, e\mu})^2 - (\pTmiss +\vec p_{T, e\mu})^2}\,, 
\label{eq:mtww1}
\end{equation}
where,
\begin{equation}
E_{\rm T, e \mu} = \sqrt{\vec p_{\rm T, e\mu}^2 +M_{e\mu}^2}\,, 
\qquad \ETmiss = \sqrt{\pTmiss^2+M_{e\mu}^2}\,.
\label{eq:mtww2}
\end{equation}
We have rescaled the LO prediction by the $K$-factor which can be read
out of table~\ref{tab:sigmacutsI} so as to be able to compare the
shapes of the distributions.  The results indicate that the shapes of
the LO distributions are mostly unchanged at NLO.

\subsection{Application to the Higgs search}
\label{sec:higgssearch}

As mentioned already, the $WW$+jet process is expected to be a
significant background to the $H(\to WW)$+jet search channel at the
LHC. In this channel there are two mechanisms by which the Higgs boson
can be produced:
\begin{itemize}
\item[(a)]
Higgs production via weak boson fusion, e.g. 
\begin{equation}
\dkp{u + d}{d + u +} {H \; ,}{W^+ W^-}  
\end{equation}
for which the characteristic signal is two jets at large rapidities
and the Higgs decay products in the central region, which contains
little jet activity.

\item[(b)]
Higgs production via gluon-gluon fusion
\begin{equation}
\dkp{g + g}{ g + } {H \; ,}{W^+ W^-} 
\label{eq:procb}
\end{equation}
where the basic interaction between the Higgs boson and the gluons is
via a top quark loop, which is usually treated in the infinite top
quark mass limit. Although the QCD partons are indicated by gluons,
the inclusion of diagrams with gluons replaced by quarks is
understood.
\end{itemize}
The NLO corrections to process (a) are
well-known~\cite{Han:1992hr,Figy:2003nv,Berger:2004pc} and small, at
the level of a few percent. In contrast, process (b) receives quite
large corrections at NLO~\cite{NLOH+1j}.  Both of these
signal processes are included at NLO in the program MCFM.

In order to make a more reliable assessment of the viability of this
search channel we will examine the effect of the inclusion of NLO
corrections to both signal and background processes when using typical
search cuts~\cite{Quayle}. These cuts are designed to enhance the WBF
process (a), whilst suppressing the large QCD backgrounds. We will be
particularly interested in examining the behaviour of the $WW$+jet
process in the search region. The background processes which we will
consider are as follows.
\begin{itemize}
\item[(c)]
QCD production of $W$-pairs plus a jet, the process for which we have
calculated the NLO corrections in this paper,
\begin{equation}
q + \bar{q} \to W^+ W^- +g \; ,
\end{equation}
where diagrams related by crossing are understood.

\item[(d)]
Associated single top production,
\begin{equation}
\dkp{g + b}{W^-+}{t \; ,}{W^+ b} 
\end{equation}
in which the top quark is produced from a $b$-quark in one of the
incoming protons. The NLO corrections to this process have been known
for some time and are typically fairly
small~\cite{Zhu:2002uj,Campbell:2005bb}. They are also implemented in
MCFM.

\item[(e)]
Top pair production,
\begin{equation}
\bothdk{g + g}{t+}{\bar{t}\; ,}{W^- \bar{b}}{W^+b } 
\end{equation}
for which the NLO corrections are known and
sizeable~\cite{Nason:1987xz}. In this study events in which two $W$'s
and only one $b$-quark are observed are considered part of the NLO
corrections to process (d). This background assumes that both
$b$-quarks are observed and it contributes because our signal contains
one or more jets. Although the corrections to the $t{\bar t}$ process
are included in MCFM, this does not include the decay of the top
quarks. Since we require cuts on the leptons produced in the top quark
decays, we therefore limit our study of this process to LO only.
 
\item[(f)]
A process analogous to the one which we consider in this paper,
$Z$-pair+jet production,
\begin{equation}
\bothdk{q + \bar{q}}{Z}{Z+g \; .}{\nu \bar{\nu}}{e^+ e^-} 
\end{equation}
The NLO corrections to this process are currently unknown and could be
computed precisely as those for $q\bar q \to WW+g$ presented in this
paper.  Accordingly, MCFM implements currently only the LO
contribution.  We will see however, that the cross-section for this
process is very small, so that no NLO corrections are required.

\item[(g)]
The production of $W$-pairs by EW processes, e.g. 
\begin{equation}
u + d \to W^- W^+ + d + u \; .
\end{equation}
This process is known at NLO~\cite{Jager:2006zc}, but is not
implemented in MCFM. Given that its cross-section turns out to be very
small, we have calculated only its leading order contribution by using
the Madgraph package~\cite{Maltoni:2002qb}.

\end{itemize} 

\begin{figure}
\begin{minipage}{.48\textwidth}
  \centering
  \includegraphics[angle=90,width=1.\textwidth]{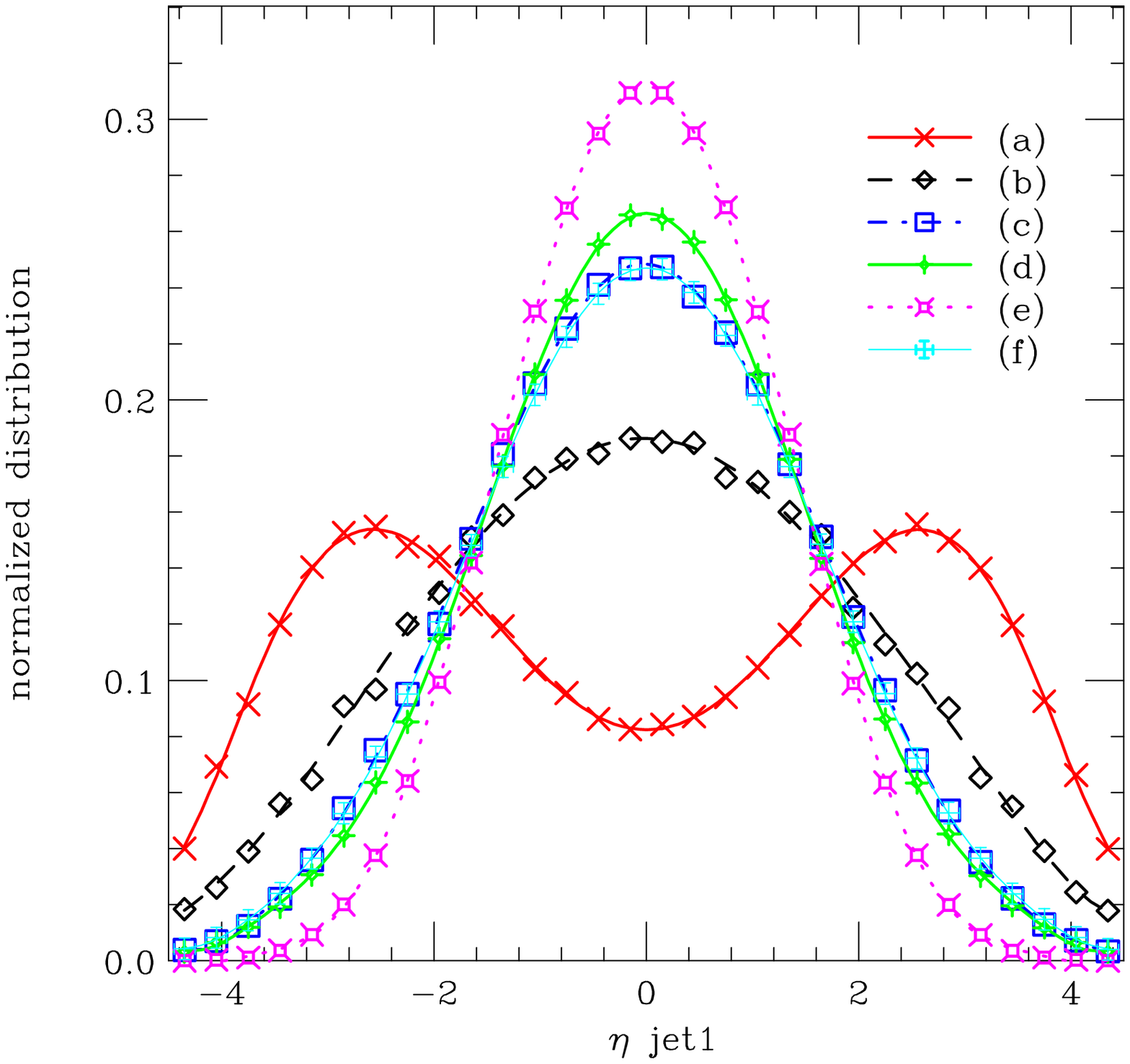}
\end{minipage}
\begin{minipage}{.48\textwidth}
\centering 
\includegraphics[angle=90,width=1.\textwidth]{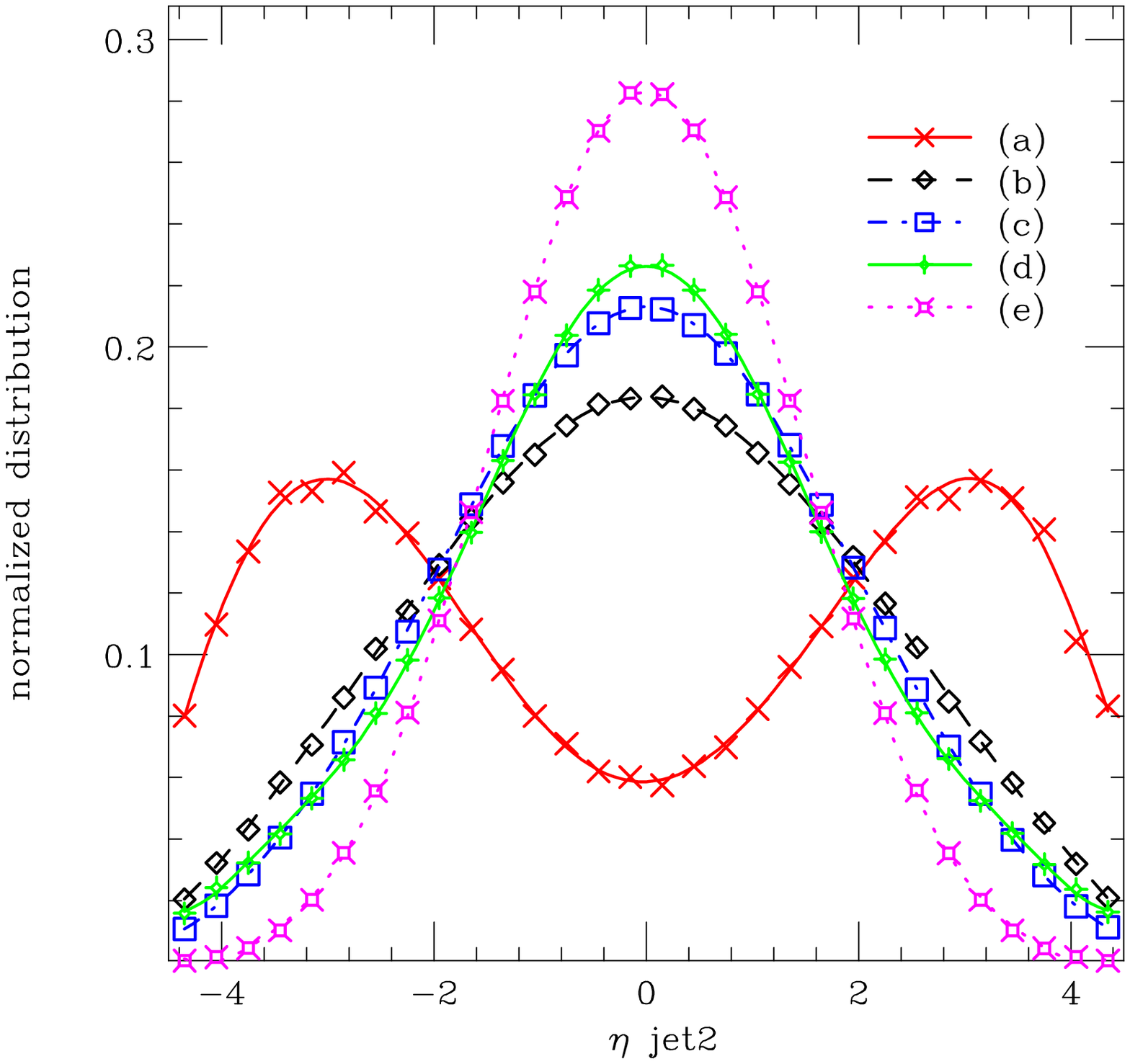}
\end{minipage}
\\
\begin{minipage}{.48\textwidth}
  \centering
  \includegraphics[angle=90,width=1.\textwidth]{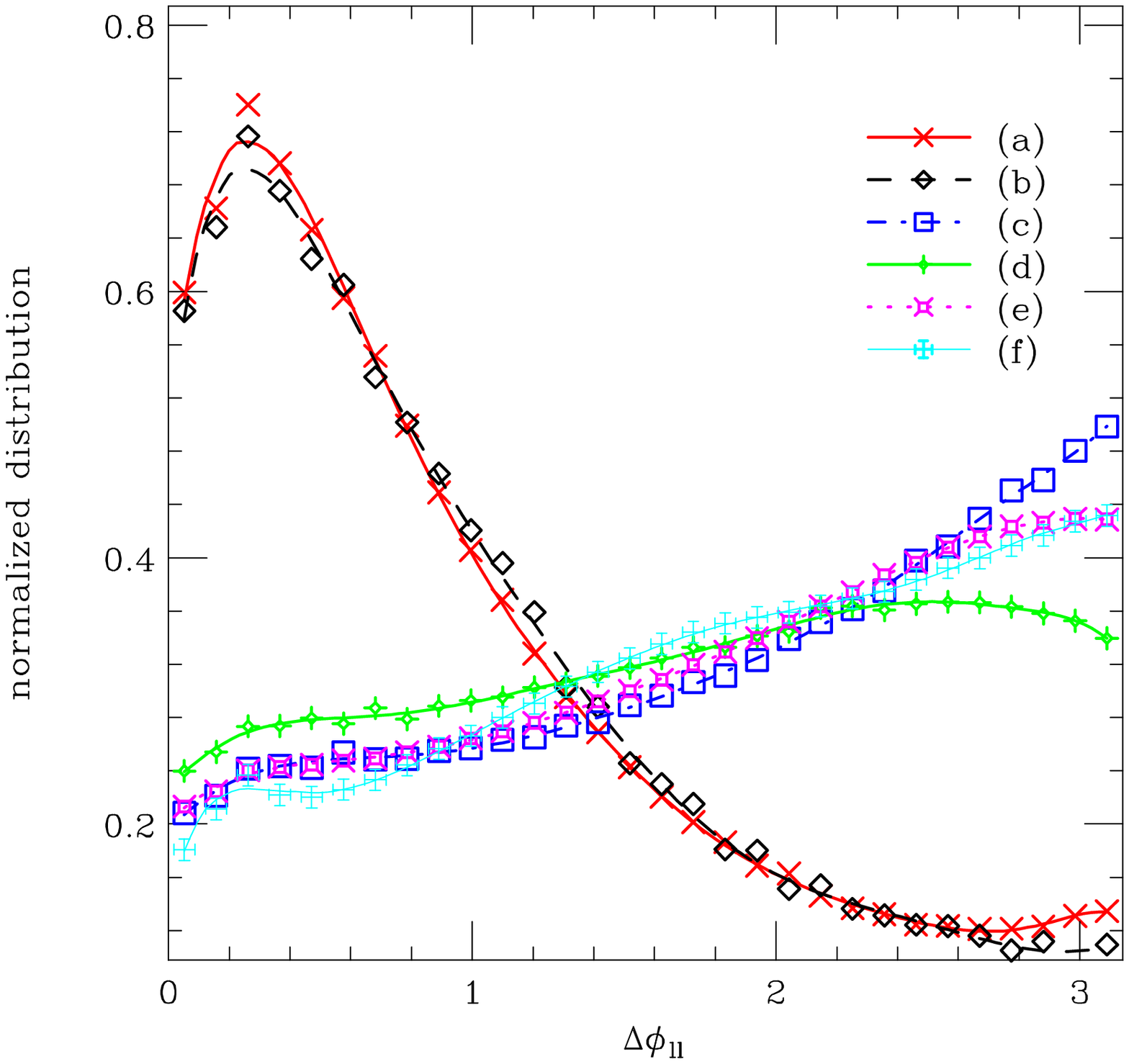}
\end{minipage}
\begin{minipage}{.48\textwidth}
\centering 
\includegraphics[angle=90,width=1.\textwidth]{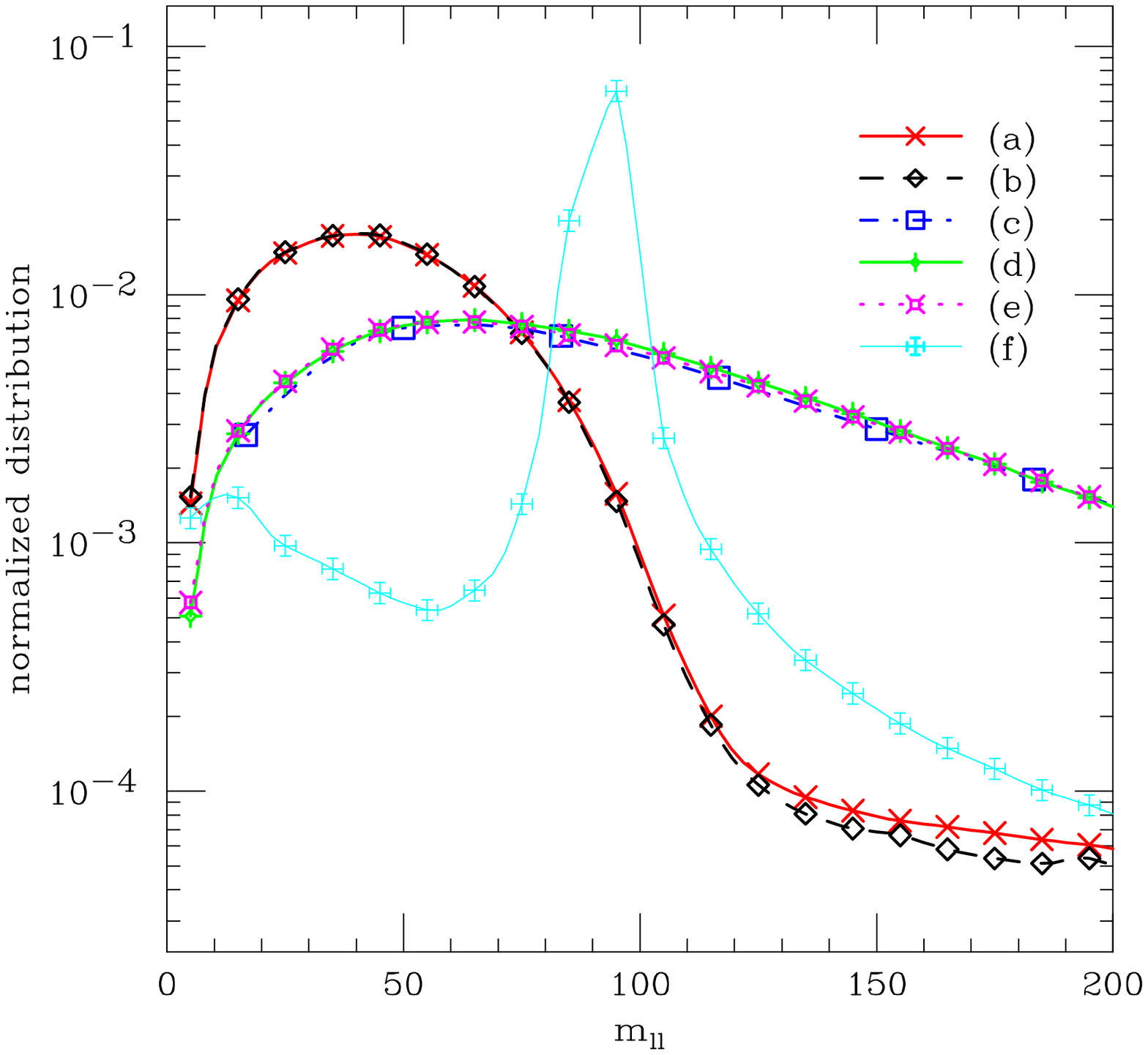}
\end{minipage}
\caption{Shapes of the signal and background distributions for
the rapidities of the leading and sub-leading jets (top-left and
top-right respectively), the angle between the two leptons
(bottom-left) and the invariant mass of the lepton pair
(bottom-right).  The curves shown are: (a) Higgs production by WBF
(red); (b) by gluon fusion (black); (c) $WW$+jet (blue); (d) $W+t$
(green); (e) $t{\bar t}$ (magenta); and
(f) $ZZ$+jet (cyan). The two Higgs production modes (by WBF and by gluon
fusion) and single top production are calculated at NLO.}
\label{fig:shapes}
\end{figure}

In order to suppress the backgrounds (c)--(g) with respect to the
Higgs boson signal processes (a) and (b), we follow the strategy of
Ref.~\cite{Quayle} and apply cuts on four additional variables. The
shapes of the distributions for both the signal and backgrounds are
shown in Figure~\ref{fig:shapes}, with only the inclusive set of cuts
(``cuts I'') applied. The signal processes are calculated using a
putative Higgs mass of $170$~GeV, above the threshold for production
of two real $W$'s.

In these figures all the curves have been normalized to unit area to
indicate only their shape, with the corresponding cross sections given
in table~\ref{tab:sigmacutsI}. We notice that with cuts I the $t \bar
t$ process is by far the dominant background.  Furthermore, since
process (g) has only a very small cross section even with this minimal
set of cuts, we do not plot the corresponding distributions in
Figure~\ref{fig:shapes} and do not consider it in the analysis any
further.
\begin{table}[h]
\caption{Cross sections (fb) for signal and background cross sections,
using cuts I only, as described in the text.}

\begin{center}
\begin{tabular}[3]{|l|c|c|}
\hline
Process & $\sigma_{LO}$ (fb) & $\sigma_{NLO}$ (fb) \\[1pt]
\hline
(a) $H \to WW$ (WBF)          & $22.5$  & $22.3$ \\
(b) $H \to WW$ (gluon fusion) & $35.9$  & $64.5$ \\
\hline
(c) $WW$+jet                  & $220$   & $269$ \\
(d) $W+t$                     & $210$   & $216$ \\
(e) $t{\bar t}$               & $3090$  & - \\
(f) $ZZ$+jet                  & $36.3$  & - \\
(g) $WW$+2 jets (EW)          & $9.3$   & - \\
\hline
\end{tabular}
\end{center}
\label{tab:sigmacutsI}
\end{table}

From these plots, it is clear that WBF production of the Higgs boson
is associated with two jets forward in rapidity.  Therefore, in
addition to our previous selection (``cuts I''), we require that the
leading jet and the second jet, if present, should be in the forward
region,
\begin{equation}
|\eta_{\rm j_1}| > 1.8 \,\qquad 
|\eta_{\rm j_2}| > 2.5 \,. 
\label{eq:jetcuts}
\end{equation}
Furthermore, the azimuthal angle between the charged leptons and their 
invariant mass should satisfy, 
\begin{equation}
\phi_{\rm \ell_1, \ell_2} < 1.2 \, \qquad m_{\rm \ell_1,\ell_2} < 75~{\rm GeV}\,. 
\label{eq:leptoncuts}
\end{equation}
The first of these cuts is a clear discriminator between the two Higgs
boson signals and all the backgrounds, whilst the second particularly
discriminates against the background from $ZZ$+jet and top quark
events.  We refer to the cuts in equations~(\ref{eq:jetcuts})
and~(\ref{eq:leptoncuts}) as ``cuts II''.

\begin{figure}
\begin{minipage}{.9\textwidth}
\centering 
\includegraphics[angle=270,width=.7\textwidth]{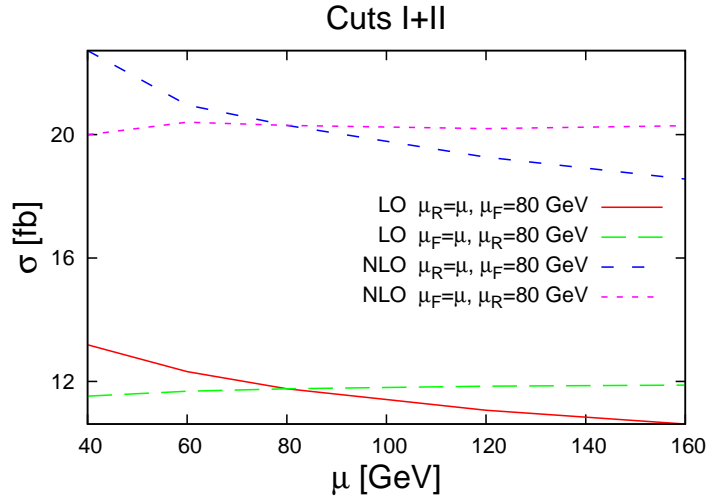}
\end{minipage}
\caption{Scale dependence for the cross section with the cuts I and II.}
\label{fig:scaledepsearch}
\end{figure}

In figure~\ref{fig:scaledepsearch} we examine the scale dependence of
the $WW$+jet background cross section with the additional cuts
applied.
There is a very large increase in the cross section going from LO to
NLO and we see very little improvement in scale dependence.  These
very large corrections are typical of calculations in which new
sub-processes enter at NLO. For $WW$+jet, the only contributions at LO
are from $q \bar q$, $q g$ and $\bar q g$ initial states. At NLO, in
the real radiation term, one has also $gg$ and $qq$ ($\bar q \bar q$)
initial states. In particular, the $qq$-type processes are very
similar kinematically to the WBF Higgs signal. Therefore when cuts II
are applied, there is a large contribution from the $qq$ initial state
since the cuts cannot suppress such ``signal-like'' sub-processes.

The cross sections for the signal and background processes, after
both cuts I and II have been applied, are collected in
table~\ref{tab:sigmacutsIandII}.
\begin{table}[h]
\caption{Cross sections (fb) for signal and background cross sections,
using cuts I+II, as described in the text.}

\begin{center}
\begin{tabular}[3]{|l|c|c|}
\hline
Process & $\sigma_{LO}$ (fb) & $\sigma_{NLO}$ (fb) \\[1pt]
\hline
(a) $H \to WW$ (WBF)          & $10.6$ & $10.6$ \\
(b) $H \to WW$ (gluon fusion) & $8.6$  & $18.0$ \\
\hline
(c) $WW$+jet                  & $11.7$   & $20.2$ \\
(d) $W+t$                     & $7.8$  & $7.6$ \\
(e) $t{\bar t}$               & $12.7$ & - \\
(f) $ZZ$+jet                  & $0.44$  & - \\
\hline
\end{tabular}
\end{center}
\label{tab:sigmacutsIandII}
\end{table}
From the cross sections given in this table we see that the cuts have
been effective in reducing the backgrounds compared to the signal
processes. Comparing with table~\ref{tab:sigmacutsI}, we see that the
WBF signal process is reduced by about a factor of $2$ and the gluon
fusion process by a factor of $3.5$. In contrast, the huge backgrounds
from top pair production and single top production have been reduced
by factors of about $300$ and $30$ respectively. The $ZZ$+jet process
has been rendered negligible as a background by the cut on the
invariant mass of the lepton pairs.  For the $WW$+jet process which we
have calculated in this paper, the cuts reduce the cross section by a
factor of $20$ at LO. However, since the NLO correction amounts now to
around $70$\%, the overall reduction of the QCD production of $WW+1$
jet at NLO is only by a factor of about $14$. As a result this process
is the dominant source of background events in our analysis, although
a naive estimate of $S/\sqrt{B}$ from this table is about $1$. We note
however that we have not included NLO effects in the estimate of the
top pair background. For the totally inclusive top cross section,
using our choice of parameters, the enhancement from LO to NLO
corresponds to a $K$-factor of approximately $1.5$.  It is therefore
reasonable to expect that the top pair cross section would also
increase at NLO in the phase space region selected by cuts I and
II. Therefore we expect the $WW$+jet and $t{\bar t}$ backgrounds to be
comparable and a significant source of background events in this
channel.

\section{Conclusions}
\label{sec:conclusions}

In this paper we have computed the NLO QCD corrections to the
production of $W$ pairs in association with a jet at hadron colliders,
a calculation highly desired by the experimental
community~\cite{Buttar:2006zd}.  The calculation is performed using a
semi-numerical approach and implemented in the general purpose NLO
code MCFM. We find that, for cuts typical of LHC analyses, in which a
jet is defined with transverse momentum above $30$~GeV, the effect of
the QCD corrections is to increase the cross section by $25$\% for our
default scale choice ($\mu_F = \mu_R = 80$~GeV). Dependence on the
renormalization and factorization scales at NLO is mild and is
decreased by about a factor of two compared to LO.

We have also performed a parton level analysis of the impact of these
QCD corrections on the search for a Higgs boson of mass $170$~GeV at
the LHC, using the channel $H \to W^+ W^- +$~jet.  Including the
effect of NLO corrections also in the signal processes, Higgs
production by gluon fusion and by WBF, we find that the $WW$+jet
background is one of the dominant backgrounds and is comparable to the
WBF signal. In this analysis we have used a tighter set of cuts to
select the WBF events and to suppress the large QCD backgrounds.  We
find that these cuts enhance the effect of the NLO corrections to
$WW$+jet, which now increase the LO cross section by $70$\% in this
region. We conclude that any studies of the Higgs boson search in this
channel must take into account this significant correction to the
number of expected background events.

\begin{acknowledgments} JC and GZ thank the Galileo Galilei Institute
for Theoretical Physics for their hospitality and the INFN for partial
support during the final stages of this work. GZ is grateful to Zuerich 
University for hospitality and the use of computer facilities. We also 
thank Babis Anastasiou, Michael Dittmar and Bill Quayle for discussions 
and we acknowledge correspondence and comparison of intermediate results 
with Gregory Sanguinetti and Stefan Karg.
\end{acknowledgments}

\appendix

\section{Triangle-Z production diagrams}
\label{app:triangles}
\begin{figure}[t]
  \centering \includegraphics[angle=270,width=12cm]{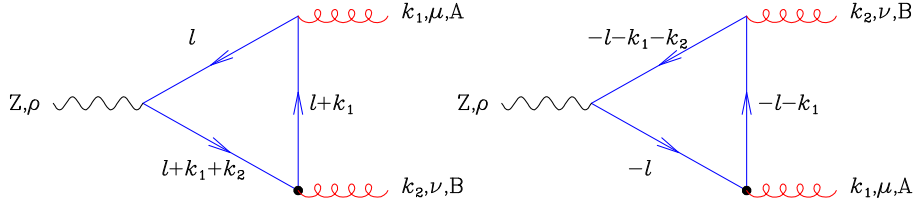}
\caption{Triangle diagrams with closed fermion loops.}
\label{fig:triangles}
\end{figure}

In this appendix we give analytical results for triangle diagrams
where a $Z$ is emitted from a closed fermion loop.  Contributions
where a Higgs is emitted are included in the signal process b),
eq.~\ref{eq:procb}.  Results are valid for an arbitrary value of the
mass $m$ of the quarks in the loop.

We calculate the triangle shown in Fig.~\ref{fig:triangles}, where all
momenta are outgoing $k_1+k_2+k_3=0$ and to begin with $k_1^2 \neq 0,
k_2^2 \neq 0$.  The result for the two triangle diagrams (including
the minus sign for a fermion loop) is,
\begin{equation} \label{EqT}
T^{\mu \nu \rho}_{AB}(k_1,k_2) = i \frac{g_s^2}{16 \pi^2} \frac{1}{2} 
\delta_{AB} 
\Big(\frac{g_W}{2 \cos \theta_W} \Big) T_3^f  \Gamma^{\mu \nu \rho}\,,
\end{equation}
where,
\begin{equation}
\Gamma^{\mu \nu \rho}(k_1,k_2,m) = \frac{2}{i \pi^2} \int \; d^n l
\; \mbox{Tr}\{ \gamma^\rho \gamma_5 \frac{1}{\slsh{l}-m}
\gamma^\mu \frac{1}{\slsh{l}+\slsh{k}_1-m}
\gamma^\nu \frac{1}{\slsh{l}+\slsh{k}_1+\slsh{k}_2-m} \}\,.
\end{equation}

The most general form of $\Gamma$ consistent with QCD gauge invariance,
\begin{equation}
k_{1}^{\mu} \Gamma_{\mu \nu \rho}=
k_{2}^{\nu} \Gamma_{\mu \nu \rho} =0 \; ,
\end{equation}
can be written as,
\begin{eqnarray}
\Gamma^{\mu \nu \rho}&=& F_1(k_1,k_2,m) \;
      \Big\{ \mbox{Tr}[\gamma^\rho \gamma^\nu \slsh{k_1} \slsh{k_2}\gamma_5 ] k_1^\mu
            +\mbox{Tr}[\gamma^\rho \gamma^\mu \gamma^\nu \slsh{k_2}\gamma_5 ] k_1^2 \Big\}\nonumber \\
  &+& F_2(k_1,k_2,m) \;
        \Big\{\mbox{Tr}[\gamma^\rho\gamma^\mu\slsh{k_1}\slsh{k_2}\gamma_5 ] k_2^\nu
        +\mbox{Tr}[\gamma^\rho\gamma^\mu\gamma^\nu\slsh{k_1}\gamma_5 ] k_2^2 \Big\}\nonumber \\
  &+& F_3(k_1,k_2,m)\; (k_1^\rho+k_2^\rho)
     \Big\{ \mbox{Tr}[ \gamma^\mu\gamma^\nu\slsh{k_1}\slsh{k_2}\gamma_5 ] \Big\}\nonumber \\
  &+& F_4(k_1,k_2,m)\; (k_1^\rho-k_2^\rho)
      \Big\{ \mbox{Tr}[\gamma^\mu\gamma^\nu\slsh{k_1}\slsh{k_2}\gamma_5 ]  \Big\}\,. 
\end{eqnarray}

The results for the coefficients $F_i$ 
are~\cite{Adler:1969gk,Hagiwara:1990dx,Hagiwara:1991xy},
\begin{eqnarray}
F_1(k_1,k_2,m) & = &  -I_{10}+I_{11}+I_{20}\,, \nonumber \\
F_2(k_1,k_2,m) & = &  +I_{01}-I_{11}-I_{02}\,, \nonumber \\
F_3(k_1,k_2,m) & = &  -I_{11}\,, \nonumber \\
F_4(k_1,k_2,m) & = &  0\,, 
\end{eqnarray}
where,
\begin{equation}
I_{st}= \int_0^1 \; dz_1 dz_2 dz_3 \; \delta(1-z_1-z_2-z_3) 
\frac{4 z_1^s z_2^t}{
[z_1 z_3 k_1^2 + z_2 z_3 k_2^2 + z_1 z_2 k_3^2-m^2]}\,, 
\end{equation}
and $k_3=(k_1+k_2)^2$.  For the particular case at hand we are
interested in $k_2^2=\varepsilon.k_2=0$, so we get a
contribution from $F_1$, and in the case of off-shell $W$'s from
$F_3$. In this limit we obtain,
\begin{eqnarray}
F_1(k_1,k_2,m) &=&\frac{1}{2 k_1 . k_2}
\Big[2+4 m^2 C_0(k_1^2,0,(k_1+k_2)^2,m^2,m^2,m^2) \nonumber \\
 &+&\Big(2 + \frac{k_1^2}{k_1 .k_2}\Big)
 \Big[B_0((k_1+k_2)^2,m^2,m^2)-B_0(k_1^2,m^2,m^2)\Big]\,,
\nonumber \\
F_3(k_1,k_2,m) &=&-F_1(k_1,k_2,m)\nonumber \\
&+&\frac{1}{k_1 . k_2} 
\Big[B_0((k_1+k_2)^2,m^2,m^2)-B_0(k_1^2,m^2,m^2)\Big]\,,
\end{eqnarray}
where,
\begin{eqnarray}
C_0(k_1^2,k_2^2,(k_1+k_2)^2,m_1^2,m_2^2,m_3^2)&=&\frac{1}{i\pi^2} \int d^dq 
\frac{1}{[q^2-m_1^2][(q+k_1)^2-m_2^2][(q+k_{12})^2-m_3^2]}\,, \nonumber \\
B_0(k_1^2,m_1^2,m_2^2)&=&\frac{1}{i\pi^2} \int d^dq 
\frac{1}{[q^2-m_1^2][(q+k_1)^2-m_2^2]}\,,
\end{eqnarray}
with $k_{12}= k_1+k_2$.  We now make the identification of the momenta
$k_1=p_1+p_2,k_3=p_3+p_4+p_5+p_6,k_2=p_7$ to calculate the physical
process.  To get the contribution to our process we have to multiply
$T^{\mu\nu\rho}$ given in Eq.~(\ref{EqT}) by the tensor $L$, dependent
on the helicity of the (12)-line and the polarization of the gluon
$\varepsilon_\nu(p_7,h_7)$ with momentum $p_7$.
A left handed quark line adds a factor,
\begin{equation}
\frac{-g_s}{s_{12}} t^A_{i_1 i_2} \langle 1-| \gamma^\mu | 2- \rangle\,. 
\end{equation}
The decay of a $Z$ splitting to a pair of $W$-bosons 
(which both decay to leptons) adds a factor,
\begin{eqnarray}
&&\frac{-g_W^3 \cos\theta_W}{2 s_{34} s_{56} (s_{127}-M_Z^2)}
\Big[ 2 g^{\rho \alpha } (p_3+p_4)^\beta 
-2 g^{\beta \rho }  (p_5+p_6)^\alpha 
+g^{\alpha \beta} (p_5+p_6-p_3-p_4)^\rho 
\Big] 
\nonumber \\
&\times & \langle 3-| \gamma_\alpha | 4- \rangle
\langle 5-| \gamma_\beta | 6- \rangle \times P_W(s_{34}) P_W(s_{56}) P_Z(s_{127})\,.
\end{eqnarray}
Collecting terms we may define $\Lambda_{\mu\nu\rho}$ as,
\begin{equation}
L^{A}_{\mu\nu\rho}(h_1,h_7) = 8 \sqrt{2} g_s g_W^3 \cos \theta_W  
(t^A)_{i_1 i_2} P_W(s_{34})P_W(s_{56}) P_Z(s_{127}) \; \Lambda_{\mu\nu\rho}(h_1,h_7)\,, 
\end{equation}
where $s_{ij}=(p_i+p_j)^2,s_{ijk}=(p_i+p_j+p_k)^2$ and,
\begin{eqnarray}
&&\Lambda_{\mu\nu\rho}(h_1,h_7) = \frac{1}{16 \sqrt{2}\, s_{12} s_{34} s_{56} 
(s_{127}-M_Z^2)} 
\langle p_1(h_1)|\gamma_\mu |p_2(h_1)\rangle \; \varepsilon_\nu(p_7,h_7) 
\nonumber \\
&\times & \langle3-|\gamma_\alpha|4-\rangle \langle5-|\gamma_\beta |6-\rangle \nonumber \\
&\times &
\Big[ 2 g^{\rho \alpha } (p_3+p_4)^\beta 
-2 g^{\beta \rho }  (p_5+p_6)^\alpha 
+g^{\alpha \beta} (p_5+p_6-p_3-p_4)^\rho 
\Big]\,. 
\end{eqnarray}

So the amplitude with a heavy doublet circulating in the loop 
is given by,
\begin{eqnarray}
{\cal A}^B_{\rm tri}(q_1^{h_1},\bar{q}_2^{\bar{h}_1},\nu_3,e^+_4,\mu^-_5,\bar{\nu}_6,g_7^{h_7}) 
&=&\sum_f T^{\mu \nu \rho}_{AB}(k_1,k_2) \times L_{\mu \nu \rho} \nonumber \\
&=& \sum_f \frac{g_s^2}{16 \pi^2} \; H \sqrt{2} \; t^B_{i_1 i_2}  \; 
\{ 2 T_3^f \} \Gamma^{\mu \nu \rho} \Lambda_{\mu \nu \rho} (h_1,h_7) \\
&=& \frac{g_s^2}{16 \pi^2} \; H \sqrt{2} t^B_{i_1 i_2}  \; 
A_{\rm tri}(q_1^{h_1},\bar{q}_2^{\bar{h}_1},\nu_3,e^+_4,\mu^-_5,\bar{\nu}_6,g_7^{h_7}) \nn\,.
\label{eq:Atridef}
\end{eqnarray}
Note that with these definitions the overall factor $H$ is the same as
in Eq.~(\ref{Overall}) except for the extra factor of $\frac{g_s^2}{16
\pi^2}$ and a sign from $2 T_3^f$ depending on whether a top or bottom
quark is circulating in the loop.

Defining ${\cal F}_i = F_i(k_1,k_2,m_t)-F_i(k_1,k_2,m_b)$
we write out in detail the case $h_1=-1,h_7=\pm 1$,
\begin{eqnarray}
\label{eq:Atri}
A_{\rm tri}(q_1^{L},\bar{q}_2^{R},\nu_3,e^+_4,\mu^-_5,\bar{\nu}_6,g_7^{R}) &=&  \sum_f \{2 T_3^f\}
\Gamma^{\mu \nu \rho} \Lambda_{\mu \nu \rho} (-1,+1) \nonumber \\ 
&=&
 \frac{-{\cal F}_1 [27]}{4 s_{34} s_{56} (s_{127}-M_Z^2) } 
\Big[
 \langle 1-|3+4-5-6|7-\rangle \langle 35 \rangle [46] \nonumber \\
 &+&2\langle 5-|3+4|6-\rangle \langle 13 \rangle [47]
 -2\langle 3-|5+6|4-\rangle \langle 15 \rangle [67] \Big]\nonumber \\
&-& \frac{{\cal F}_3  [27]^2}{4 s_{12} s_{34} s_{56} (s_{127}-M_Z^2) }  
\langle 12\rangle (s_{56}-s_{34}) \langle 35 \rangle [46]\,,
\end{eqnarray}
\begin{eqnarray}
A_{\rm tri}(q_1^{L},\bar{q}_2^{R},\nu_3,e^+_4,\mu^-_5,\bar{\nu}_6,g_7^{L}) &=& 
\sum_f \{2 T_3^f\}
\Gamma^{\mu \nu \rho} \Lambda_{\mu \nu \rho} (-1,-1)\nonumber \\ 
&=& 
\frac{-{\cal F}_1  \langle 17\rangle  }{4 s_{34} s_{56} (s_{127}-M_Z^2)} 
\Big[
  \langle 7-|3+4-5-6|2-\rangle \langle 35 \rangle [46] \nonumber \\
 &+&2\langle 3-|5+6|4-\rangle \langle 75 \rangle [26]
 -2\langle 5-|3+4|6-\rangle \langle 73 \rangle [24]
\Big]\nonumber \\
&-& \frac{{\cal F}_3  \langle 17\rangle^2}
{4 s_{12} s_{34} s_{56} (s_{127}-M_Z^2)} [12](s_{34}-s_{56}) 
\langle 35 \rangle [46]\,. 
\end{eqnarray}
Switching between the two gluon helicities is given by $-$flip$_1$,
with the flip$_1$ symmetry defined as,
\begin{equation}
\mbox{flip}_1: \;\; 1 \leftrightarrow 2, \;\;
3 \leftrightarrow 6, \;\;4 \leftrightarrow 5, \;\;\langle ab\rangle  \leftrightarrow [ab] \; .
\end{equation}

The cases $h_1=+1,h_7=\pm 1$ are simply given by, 
\begin{equation}
A_{\rm tri}(q_1^{R},\bar{q}_2^{L},\nu_3,e^+_4,\mu^-_5,\bar{\nu}_6,g_7^{L/R}) = 
A_{\rm tri}(q_2^{L},\bar{q}_1^{R},\nu_3,e^+_4,\mu^-_5,\bar{\nu}_6,g_7^{L/R})\,. 
\end{equation}

Similarly to Eq.~(\ref{eq:Atridef}) we define the contribution from
the six massive box diagrams as,
\begin{equation}
\label{eq:Abox}
{\cal A}^{B}_{\rm box}(q_1^{h_1},\bar{q}_2^{\bar h_1},\nu_3,e^+_4,\mu^-_5,\bar{\nu}_6,g_7^{h_7})
= \frac{g_s^2}{16 \pi^2} H \; \sqrt{2} \; t^B_{i_1 i_2} \;  A_{\rm box}(q_1^{h_1},\bar{q}_2^{\bar h_1},\nu_3,e^+_4,\mu^-_5,\bar{\nu}_6,g_7^{h_7})\,. 
\end{equation}
In the next appendix we give numerical results for $A_{\rm
  tri}(q_1^{h_1},\bar{q}_2^{\bar h_1},\nu_3,e^+_4,\mu^-_5,\bar{\nu}_6,g_7^{h_7})$
and $A_{\rm
  box}(q_1^{h_1},\bar{q}_2^{\bar h_1},\nu_3,e^+_4,\mu^-_5,\bar{\nu}_6,g_7^{h_7})$
for a randomly chosen phase space point.

\section{Explicit numerical results}
\label{app:numres}
In this section we present explicit numerical results for one phase
space point for the independent amplitudes entering the one-loop cross
sections.

We define the one-loop virtual amplitudes analogously to
Eqs.(\ref{eq:treeAu}, \ref{eq:treeAd}, \ref{eq:Atri} and
\ref{eq:Abox}), e.g.
\begin{eqnarray}
\cA^{A}_v(u^L_1,\bar{u}^R_2,\nu_3,e^+_4,\mu^-_5,\bar{\nu}_6,g_7^R)&=& 
\frac{g_s^2}{16\pi^2} \sqrt{2}\, t^A _{i_1 i_2}\, H
%
\\
&&\Bigg[A_{7v}^{(a)}(1,2,3,4,5,6,7)+C_{L,\{u\}}(s_{127}) A_{7v}^{(b)}(1,2,3,4,5,6,7)\Bigg]\nonumber\,, 
\label{eq:oneloopALu}
\end{eqnarray}
and we have similar definitions for the other amplitudes. 

In table~\ref{table:results} we present results for the colour
stripped virtual amplitudes $A_{7v}^{(a)}(1,2,3,4,5,6,7)$ and
$A_{7v}^{(b)}(1,2,3,4,5,6,7)$ for positive and negative quark and
gluon polarizations at the following randomly chosen phase space point
at the LHC -- $(p_x,p_y,p_z,E)~[\mbox{GeV}]$ (all momenta are
outgoing) at $\mu_R = 80$~GeV:
\begin{displaymath}
\begin{array}{rrrrrrr}
     p_1 & = &\!(&\! 0.00000000000000,&\! 0.00000000000000,&\! 1021.22119318758,&\!-1021.22119318758),\\
     p_2 & = &\!(&\! 0.00000000000000,&\! 0.00000000000000,&\!-238.714576090637,&\!-238.714576090637),\\
     p_3 & = &\!(&\!-71.5344542606618,&\!-183.877222508616,&\!-3.11006048502754,&\! 197.326337775966),\\
     p_4 & = &\!(&\!-9.92033815503652,&\!-76.1125676676337,&\! 49.0057636944973,&\! 91.0664644166627),\\
     p_5 & = &\!(&\! 32.5059044554765,&\! 245.099246845329,&\!-495.644737899924,&\! 553.889863453468),\\
     p_6 & = &\!(&\! 64.1786550096635,&\! 124.613643938661,&\!-207.896850811885,&\! 250.736037681104),\\
     p_7 & = &\!(&\!-15.2297670494417,&\!-109.723100607740,&\!-124.860731594604,&\! 166.917065951017).\\
\end{array}
\end{displaymath}

\begin{scriptsize}
\TABLE{
\begin{tabular}{|c|c|c|c|}
\hline
Amplitude &  $c_\Gamma/\epsilon^2 \alpha_s/(4\pi) $ & $c_\Gamma/\epsilon \alpha_s/(4\pi)$ & $\alpha_s/(4\pi)$\\
\hline
\hline
$|A_7^a(1^R,2^L,3,4,5,6,7^L)| $    & - & - & $ 0 $ \\
\hline
$|A_{v7}^a(1^R,2^L,3,4,5,6,7^L)| $ & $ 0 $ & $ 0.413542\cdot10^{-16} $ & $ 0.116250\cdot10^{-6}$ \\
\hline
$|A_7^b(1^R,2^L,3,4,5,6,7^L)| $    & - & - & $ 0.209219\cdot10^{-4} $ \\
\hline
$|A_{v7}^b(1^R,2^L,3,4,5,6,7^L)| $ & $ 0.118557\cdot10^{-3} $ & $ 0.238602\cdot10^{-3} $ & $ 0.473769\cdot10^{-3}$ \\
\hline
$|A_7^a(1^L,2^R,3,4,5,6,7^L)| $    & - & - & $ 0.695412\cdot10^{-5} $ \\
\hline
$|A_{v7}^a(1^L,2^R,3,4,5,6,7^L)| $ & $ 0.394067\cdot10^{-4} $ & $ 0.793076\cdot10^{-4} $ & $ 0.127832\cdot10^{-3}$ \\
\hline
$|A_7^b(1^L,2^R,3,4,5,6,7^L)| $    & - & -& $ 0.686719\cdot10^{-5} $ \\ 
\hline
$|A_{v7}^b(1^L,2^R,3,4,5,6,7^L)| $ & $ 0.389141\cdot10^{-4} $ & $ 0.783161\cdot10^{-4} $ & $ 0.125011\cdot10^{-3}$ \\
\hline
$|A_7^a(1^R,2^L,3,4,5,6,7^R)| $    & - & - & $ 0 $ \\
\hline
$|A_{v7}^a(1^R,2^L,3,4,5,6,7^R)| $ & $ 0  $ & $ 0.274681\cdot10^{-16} $ & $ 0.216005\cdot10^{-6}$ \\
\hline
$|A_7^b(1^R,2^L,3,4,5,6,7^R)| $    & - & - & $ 0.171644\cdot10^{-4} $ \\ 
\hline
$|A_{v7}^b(1^R,2^L,3,4,5,6,7^R)| $ & $ 0.972651\cdot10^{-4} $ & $ 0.195750\cdot10^{-3} $ & $ 0.352339\cdot10^{-3}$ \\
\hline
$|A_7^a(1^L,2^R,3,4,5,6,7^R)| $    & - & - & $ 0.104711\cdot10^{-4} $ \\
\hline
$|A_{v7}^a(1^L,2^R,3,4,5,6,7^R)| $ & $ 0.593364\cdot10^{-4} $ & $ 0.119417\cdot10^{-3} $ & $ 0.254739\cdot10^{-3}$ \\
\hline
$|A_7^b(1^L,2^R,3,4,5,6,7^R)| $    & - & - & $ 0.102672\cdot10^{-4} $ \\
\hline
$|A_{v7}^b(1^L,2^R,3,4,5,6,7^R)| $ & $ 0.581805\cdot10^{-4} $ & $ 0.117091\cdot10^{-3} $ & $ 0.247373\cdot10^{-3}$ \\
\hline
\end{tabular}
\caption{Numerical results for the Born and virtual amplitudes with massless particles in the loop.}
\label{table:results}
}
\end{scriptsize}

\begin{table}
\begin{scriptsize}
\begin{center}
\begin{tabular}{|c|c|}
\hline
Amplitude & $\alpha_s/(4\pi)$\\
\hline
\hline
$|A_{\rm box}(1^R,2^L,3,4,5,6,7^L)| $ & $\,2.96740 10^{-8}$ \\
$|A_{\rm tri}(1^R,2^L,3,4,5,6,7^L)| $ & $\,8.48025 10^{-7}$ \\
$|A_{\rm box}(1^L,2^R,3,4,5,6,7^L)| $ & $\,4.83946 10^{-8}$ \\
$|A_{\rm tri}(1^L,2^R,3,4,5,6,7^L)| $ & $\,1.20074 10^{-7}$ \\
$|A_{\rm box}(1^R,2^L,3,4,5,6,7^R)| $ & $\,7.35610 10^{-8}$ \\
$|A_{\rm tri}(1^R,2^L,3,4,5,6,7^R)| $ & $\,1.13325 10^{-6}$ \\
$|A_{\rm box}(1^L,2^R,3,4,5,6,7^R)| $ & $\,1.74823  10^{-8}$ \\
$|A_{\rm tri}(1^L,2^R,3,4,5,6,7^R)| $ & $\,1.14869 10^{-7}$ \\
\hline
\end{tabular}
\end{center}
\caption{Numerical results for the massive triangle and box contributions to the virtual amplitude.}
\label{table:massiveresults}
\end{scriptsize}
\end{table}

The poles of the one-loop unrenormalized virtual amplitudes are given
by
\begin{eqnarray}
A_{7v}^{a/b}(1,2,3,4,5,6,7) &=& A_{7}^{a/b}(1,2,3,4,5,6,7) 
\frac{\alpha_s}{4\pi}c_\Gamma \mu_R^{2\epsilon}\left(-\frac{3
C_F}{\epsilon}\right. \\
&+&\left.
\frac{1}{\epsilon^2}
\left(N_c (-s_{17})^{-\epsilon}+N_c(-s_{17})^{-\epsilon}
-\frac{1}{N_c}(-s_{12})^{-\epsilon}\right)
\right)\,.
\end{eqnarray}
We renormalize the amplitude by adding the $\MSbar$ counterterm 
\begin{equation}
A_{7, \rm ct}(1,2,3,4,5,6,7) = -\frac{\cG}{\epsilon} b_0 \frac{\alpha_s}{8 \pi} A_7(1,2,3,4,5,6,7)\,, 
\end{equation}
where 
\begin{equation}
b_0 = \frac{11 N_c - 4 n_f T_R}{3}\,,\qquad \mbox{and}\qquad 
\cG = (4\pi)^\epsilon \frac{\Gamma(1+\epsilon)\Gamma^2(1-\epsilon)}{\Gamma(1-2\epsilon)} \,. 
\end{equation}
In table~\ref{table:massiveresults} we give numerical results
for the finite contributions to the amplitude coming from the massive
triangles and boxes.

\end{document}